\documentclass[aps,amsmath,amssymb,nofootinbib,twocolumn]{revtex4}
\usepackage{graphicx}
\usepackage{pgfplots}
\pgfplotsset{compat=1.17}
\usepackage{dcolumn}
\usepackage{color}
\usepackage{scrextend}
\usepackage{subfig}
\usepackage{bm}
\usepackage{amsmath,amssymb,graphicx}
\usepackage{epsfig}
\usepackage{enumerate}
\usepackage{array}
\usepackage{multirow}
\usepackage{tikz}
\usepackage{ragged2e}
\usepackage{textgreek}
\usepackage[normalem]{ulem}

\begin{document}
\title{Quantum jumps in open cavity optomechanics and\\ Liouvillian versus Hamiltonian exceptional points}
\author{Aritra Ghosh\footnote{aritraghosh500@gmail.com} and M. Bhattacharya}
\affiliation{School of Physics and Astronomy, Rochester Institute of Technology, 84 Lomb Memorial Drive, Rochester, New York 14623, USA}
\vskip-2.8cm
\date{\today}
\vskip-0.9cm

\vspace{5mm}
\begin{abstract}
Exceptional points, where two or more eigenstates of a non-Hermitian system coalesce, are now of interest across many fields of physics, from the perspective of open-system dynamics, sensing, nonreciprocal transport, and topological phase transitions. In this work, we investigate exceptional points in cavity optomechanics, a platform of interest to diverse communities working on gravitational-wave detection, macroscopic quantum mechanics, quantum transduction, etc. Specifically, we clarify the role of quantum jumps in making a clear distinction between Liouvillian and Hamiltonian exceptional points in optomechanical systems. While the Liouvillian exceptional point arises from the unconditional Lindblad dynamics and is independent of the phonon-bath temperature, the Hamiltonian exceptional point emerges from the conditional no-jump evolution and acquires a thermal shift due to an enhanced conditional damping. Employing the thermofield formalism, we derive a unified spectral framework that interpolates between these regimes via an analytical hybrid-Liouvillian description. Remarkably, in the weak-quantum-jump regime, the exceptional point is perturbed only at the second order, highlighting the robustness of the Hamiltonian exceptional point under small hybrid perturbations. Our work reveals a continuous family of hybrid exceptional points, clarifies the operational and physical differences between the conditional and unconditional dissipative dynamics in optomechanical systems, and provides a probe for thermal baths.
\end{abstract}

\maketitle

\section{Introduction}
In recent years, there has been a significant interest in understanding non-Hermitian systems \cite{Moiseyev_2011,Bagarello_2015}, motivated by the discovery of $\mathcal{PT}$-symmetry \cite{Bender_1998,Bender_1999,Ruter_2010,ElGanainy_2018}, and also due to the interest in open quantum systems which evolve in a nonunitary fashion \cite{Rotter_2009}. Non-Hermitian systems are associated with the unique presence of exceptional points or non-Hermitian degeneracies where two or more eigenstates (eigenvalues and eigenvectors) coalesce \cite{Berry_2004,Heiss_2012}. Such singular points have been experimentally observed \cite{Kim_2016,Liang_2023} and have attracted a lot of attention due to potential applications in lasing \cite{Peng_2014,Zhang_2022}, sensing \cite{Lau_2018,Wiersig_2020}, mode switching due to dynamical encircling \cite{Doppler_2016}, etc. Among the available platforms for studying non-Hermitian phenomena, an important testbed is provided by cavity optomechanics \cite{Zhang_2022,Xiong_2021,Pino_2022,Sun_2023,Wang_2025} where the photon and phonon baths typically act as sources of dissipation, i.e., non-Hermiticity. The corresponding exceptional points arise as singularities in the spectrum of the Liouvillian superoperator that dictates the nonunitary evolution of the density matrix and are dubbed Liouvillian exceptional points (LEPs) \cite{Minganti_2019,Minganti_2020,Khandelwal_2021,Perina_2022,Chimczak_2023,Sun_2024,Abo_2024,Kopciuch_2025}.

\vspace{2mm}

Concerning open quantum systems with Markovian dissipation, the time evolution is dictated by the Lindblad equation \cite{Lindblad_1976,Gorini_1976}, typically obtained under the Born-Markov and rotating-wave approximations. For the conditional dynamics in the absence of quantum jumps, the system's evolution mimics that generated by an effective non-Hermitian Hamiltonian $H_{\rm NH} = H - i \Sigma$, where $H$ dictates the coherent or conservative dynamics while $\Sigma \succeq 0 $ is positive-semidefinite and describes the dissipative evolution \cite{Dalibard_1992}. One can now ask whether the associated exceptional points (if any) coincide with the LEPs. The answer, made clear in some recent works \cite{Minganti_2019,Minganti_2020,Chimczak_2023,Kopciuch_2025}, is that the exceptional points associated with $H_{\rm NH}$ are generally distinct from the LEPs. This new kind of exceptional points arising in the conditional no-jump evolution has been termed Hamiltonian exceptional points (HEPs). 

\vspace{2mm}

In this paper, we shall address the critical distinction between LEPs and HEPs in cavity optomechanics. Although exceptional points are frequently identified through effective non-Hermitian Hamiltonians, open quantum systems often require a Liouvillian description in which quantum jumps and thermal fluctuations play a fundamental role. Even though the mathematical distinction between LEPs and HEPs has been noted in the recent literature \cite{Minganti_2019,Minganti_2020,Kopciuch_2025}, its manifestation in mesoscopic systems with asymmetric environments, i.e., the zero-temperature optical cavity and the finite-temperature phonon bath inherent to optomechanics, remains an unexplored frontier. Understanding how these different spectral notions emerge and relate to one another is essential for interpreting experiments on realistic dissipative platforms. 

\vspace{2mm}

To bridge this gap and provide a transparent interpolation between LEPs and HEPs following \cite{Minganti_2020,Kopciuch_2025}, we shall make use of the thermofield formalism \cite{Takahashi_1996} which hinges on mapping a thermal-state density matrix $\rho$ onto a pure-state vector $|\rho \rangle$ via the vectorization \cite{Takahashi_1996,Jamiolkowski_1972,Choi_1975} $|\rho\rangle = \sum_{ij} \rho_{ij} |i\rangle \otimes |\tilde{j}\rangle$. As a result, the Lindblad equation gets converted to a Schr\"odinger-type equation with a non-Hermitian Hamiltonian, thereby allowing a straightforward non-Hermitian-Hamiltonian-based description of the full (unconditional) dissipative dynamics, of which the conditional no-jump dynamics is only a special case. We will apply this formalism to the red-sideband regime of good-cavity optomechanics, critical to cooling, sensing, and information storage and retrieval \cite{Aspelmeyer_2014}. Our analysis will explicitly unravel how the interplay between no/partial quantum jumps and the thermal nature of the phonon bath fundamentally leads to HEPs and hybrid exceptional points that are distinct from the LEPs, thereby offering a comprehensive spectral framework for interpreting future optomechanics experiments. We should emphasize here that the `hybrid' exceptional points of this work arise from the hybrid-Liouvillian description of \cite{Minganti_2020} and should not be confused with hybrid diabolical exceptional points \cite{Perina_2022}.

\vspace{2mm}

The paper is organized as follows. In Sec. (\ref{liouville_sec}), considering a generic optomechanical system and focusing on the red-sideband physics, we shall analyze the drift matrix for the Liouvillian (unconditional) dynamics, exposing the LEP, and will present the correlation functions. Then in Sec. (\ref{LvsH_sec}), we will determine the HEP that arises from the conditional no-jump evolution and contrast it with the LEP found earlier, exposing the role of thermal phonons leading to enhanced conditional damping. Sec. (\ref{hybrid_sec}) will then be devoted to the study of hybrid exceptional points that interpolate between the Liouvillian and Hamiltonian regimes by working in the full thermofield space. We will finally conclude the paper in Sec. (\ref{conc_sec}). Some expository details are relegated to the Appendices (\ref{appA}), (\ref{appB}), and (\ref{appC}). 

\section{Liouvillian dynamics}\label{liouville_sec}
\begin{figure}
    \centering
    \includegraphics[width=1\linewidth]{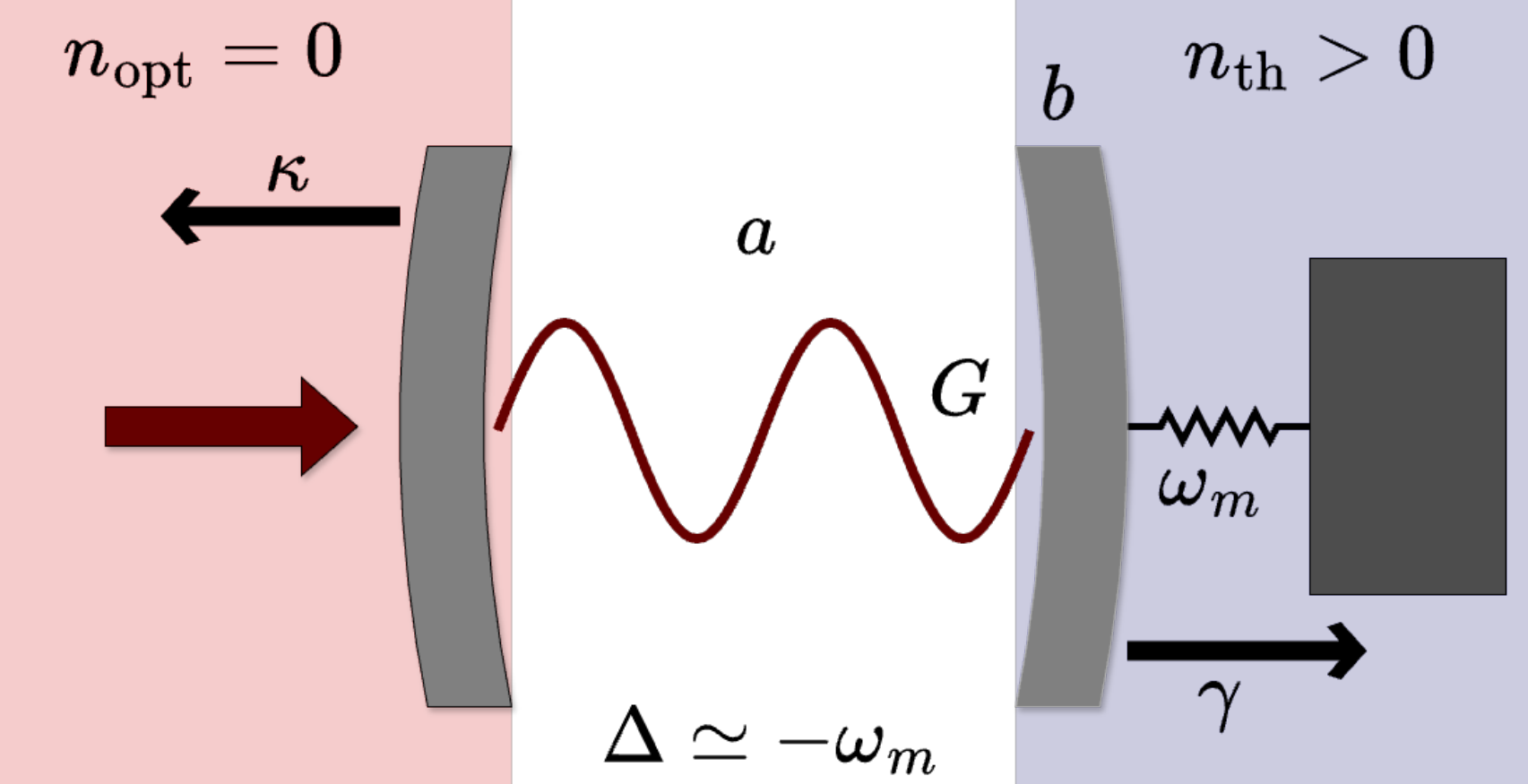}
    \caption{\justifying{Schematic of an optomechanical setup where a control laser enters the cavity from the left mirror and the intracavity mode $a$ couples with the mechanical motion of the right mirror, represented by the oscillator-mode $b$. 
    The mechanical eigenfrequency is $\omega_m$ and $G$ is the effective (linearized) optomechanical coupling. The optical and mechanical decay rates are $\kappa$ and $\gamma$, respectively. We shall take $\omega_m \gg \kappa, G, \gamma$ in conformity with the good-cavity requirement for which taking the cavity detuning $\Delta \simeq -\omega_m$ allows us to write a beam-splitter form of the linearized Hamiltonian. The photon and phonon baths are shaded with red and blue, respectively.}}
\label{fig0}
\end{figure}

Our starting point is the linearized optomechanical Hamiltonian \cite{Kippenberg_2007,Aspelmeyer_2014} in the good-cavity regime and near the red sideband $\Delta \simeq - \omega_m$ (see Fig. (\ref{fig0}) for details), given by (taking $\hbar =1$)
\begin{equation}
H = -\Delta a^\dagger a + \omega_m b^\dagger b + G a^\dagger b + G ab^\dagger,
\end{equation}
where the bosonic operators $(a,a^\dagger)$ and $(b,b^\dagger)$ correspond to the photon and the phonon degrees of freedom, respectively, while $\Delta$, $\omega_m$, and $G$ denote the cavity detuning, mechanical eigenfrequency, and linearized optomechanical coupling, respectively. Here the (strong) classical drive that pumps the cavity is absorbed into the steady-state intracavity amplitude, thereby yielding the effective optomechanical coupling $G \propto \sqrt{P_{\rm in}}$, where $P_{\rm in}$ is the input power \cite{Aspelmeyer_2014}.
Taking dissipation of the optical and mechanical modes into account, the open-system dynamics is described by the Lindblad equation
\begin{equation}
\frac{d\rho}{dt} = -i[H,\rho] + \kappa \mathcal{D}[a]\rho + \gamma (n_{\rm th}+1)\mathcal{D}[b]\rho + \gamma n_{\rm th}\mathcal{D}[b^\dagger]\rho,
\end{equation}
with the dissipator $\mathcal{D}[o]\rho = o\rho o^\dagger - \frac{1}{2}(o^\dagger o\rho + \rho o^\dagger o)$. Here $\kappa$ is the rate of cavity damping and its mechanical counterpart is $\gamma$, with $\kappa \gg \gamma$. We have taken the optical bath to be at zero temperature. This is well justified since typical optical frequencies ($\omega_{\rm opt}/2\pi \sim 10^{14}~{\rm Hz}$) are orders-of-magnitude larger than the thermal scale $k_B T/\hbar$ even at room temperature, leading to a vanishingly-small photon occupation, i.e., $n_{\rm opt} \simeq 0$, so no term containing $\mathcal{D}[a^\dagger]$ appears. In contrast, the mechanical or phonon bath is at nonzero temperature and with thermal occupation $n_{\rm th} > 0$. Now the dynamics of an operator $O$ follows from the adjoint Lindblad evolution $\frac{dO}{dt} = \mathcal{L}^\dagger (O)$, given by
\begin{eqnarray}
\frac{dO}{dt}&=& i[H,O]+\kappa\left(a^\dagger O a-\frac{1}{2}\{a^\dagger a,O\}\right) \\
&&+\gamma(n_{\rm th}+1)\left(b^\dagger O b-\frac{1}{2}\{b^\dagger b,O\}\right) \nonumber \\
&&+\gamma n_{\rm th}\left(b O b^\dagger-\frac{1}{2}\{b b^\dagger,O\}\right), \nonumber
\end{eqnarray}
and is equivalent to the standard quantum Langevin form upon adding input-noise operators \cite{Gardiner_1985}. Dropping the noise terms but keeping the deterministic drift yields
\begin{eqnarray}
\frac{da}{dt} &=& \left(i\Delta-\frac{\kappa}{2}\right)a - iG b, \label{Ha} \\
\frac{db}{dt} &=& \left(-i\omega_m-\frac{\gamma}{2}\right)b - iG a, \label{Hb}
\end{eqnarray}
together with the Hermitian-conjugate equations. The bath occupation $n_{\rm th}$ enters only through the noise correlations (e.g., $\langle b_{\rm in}^\dagger(t)b_{\rm in}(t')\rangle=n_{\rm th}\delta(t-t')$),
not through the drift matrix. In order to cast these into the matrix form which will facilitate the identification of the coefficient (drift) matrix, let us define $\Psi=
(a, ~b)^T$, so the Heisenberg equations can be expressed as 
\begin{equation}
\frac{d\Psi}{dt} = M_{ab} \Psi, \quad \quad {M}_{ab}=
\begin{pmatrix}
i\Delta-\frac{\kappa}{2} & -iG\\ 
-iG & -i\omega_m-\frac{\gamma}{2}
\end{pmatrix}.
\end{equation} 
Denoting by $\lambda_\pm$ the eigenvalues of $M_{ab}$, straightforward algebra gives
\begin{equation}\label{lambda_pm}
\lambda_{\pm} = \frac{1}{2}\Big(i(\Delta-\omega_m)-\frac{\kappa+\gamma}{2}\Big) \pm \sqrt{D},
\end{equation}
where 
\begin{equation}
D = \frac{1}{4}\Big(i(\Delta+\omega_m)-\frac{\kappa-\gamma}{2}\Big)^2 -G^2.
\end{equation}
The eigenvalues coalesce when $D=0$. Enforcing $\Delta = -\omega_m$ implies that this happens when 
\begin{equation}\label{G_EP}
G_{\rm LEP} = \frac{\kappa - \gamma}{4},
\end{equation} because $\kappa > \gamma$. Now the coalesced eigenvalue is $\lambda_{\rm LEP} = i\Delta - (\kappa + \gamma)/4$ and this implies
\begin{equation}
M_{ab} - \lambda_{\rm LEP} \mathbb{I} = \frac{\kappa - \gamma}{4} 
\begin{pmatrix}
-1 & -i\\ 
-i & 1
\end{pmatrix},
\end{equation}
which has rank one, not zero. In other words, this corresponds to an exceptional point of second order and the subscript LEP is to indicate that this exceptional point is a singularity of the Liouvillian dynamics. By contrast, in the blue-sideband rotating-wave regime, the light-matter interaction is of the two-mode-squeezing type $\sim (a^\dagger b^\dagger + ab)$, and one can check that the corresponding drift block has the eigenvalue splitting $\sqrt{(\kappa-\gamma)^2/16+G^2}$. This does not vanish at any nonzero and real $G$, excluding the possibility of an exceptional point.

\vspace{2mm}

It should be clarified here that while an LEP should arise as an exceptional point in the spectrum of the full Liouvillian $\mathcal{L}$ and not just the drift matrix $M_{ab}$, the Gaussian nature of the Lindblad evolution for linearized optomechanics implies that $M_{ab}$ alone controls the relevant physics of the first moments and the two-point correlations. In other words, the coalescence condition $D=0$ for $M_{ab}$ coincides with the LEP that governs the relevant relaxation and hybrid-mode dynamics in the Gaussian sector (assuming stability, i.e., $\mathrm{Re}[\lambda_\pm]<0$). It may be strongly emphasized that the use of $M_{ab}$ does not amount to discarding quantum jumps. Rather, $M_{ab}$ is obtained from the adjoint Lindblad equation applied to the closed linear operator sector $\Psi=(a,b)^T$. The dissipators, including their jump and anti-commutator pieces, explicitly generate the damping terms in the drift equation.

\vspace{2mm}

The steady-state correlation functions can be easily calculated using the quantum regression theorem \cite{Lax_1963}, leading to the matrix form (for $\tau = t - t' \geq 0$)
\begin{equation}
C(\tau) = 
\begin{pmatrix}
\langle a^\dagger(t)a(t')\rangle_{\rm ss}
&
\langle a^\dagger(t)b(t')\rangle_{\rm ss}
\\[6pt]
\langle b^\dagger(t)a(t')\rangle_{\rm ss}
&
\langle b^\dagger(t)b(t')\rangle_{\rm ss}
\end{pmatrix}
=
e^{M_{ab}^*\tau} \mathcal{V},
\label{Ctau_matrix_form}
\end{equation}
where $\mathcal{V}$ is the equal-time steady-state covariance matrix
\begin{equation}
\mathcal{V}=
\begin{pmatrix}
n_a & \mathcal{C}_{ab}\\
\mathcal{C}_{ba} & n_b
\end{pmatrix},
\label{V_def}
\end{equation}
with $n_a=\langle a^\dagger a\rangle_{\rm ss}$, $n_b=\langle b^\dagger b\rangle_{\rm ss}$, $\mathcal{C}_{ba}=\langle b^\dagger a\rangle_{\rm ss}$, and $\mathcal{C}_{ab}=\mathcal{C}_{ba}^*$. Note that $M_{ab}^*$ denotes the element-wise complex conjugate of $M_{ab}$, not the Hermitian adjoint $M_{ab}^\dagger$. Since $M_{ab}^*$ is a $2\times2$ matrix, its exponential admits the well-known spectral expansion
\begin{equation}
e^{M_{ab}^*\tau} = P_+^*e^{\lambda_+^*\tau} + P_-^*e^{\lambda_-^*\tau},
\quad \quad
P_\pm^*=\frac{M_{ab}^*-\lambda_\mp^* \mathbb{I}}{\lambda_\pm^*-\lambda_\mp^*}.
\label{spectral_expansion}
\end{equation}
Substituting into Eq. (\ref{Ctau_matrix_form}), each correlation function is a sum of two damped exponentials. Invoking $\Delta = -\omega_m$, one finds the following closed-form expressions (see Appendix (\ref{appA})): 
\begin{eqnarray}
\langle a^\dagger(t)a(t')\rangle_{\rm ss} &=&
\sum_{\sigma=\pm} e^{\lambda^*_\sigma \tau} \left[\frac{1}{2}\left(1-\sigma\frac{\Gamma}{\delta}\right)n_a
+\sigma\frac{iG}{2\delta} \mathcal{C}_{ba}
\right], \nonumber \\
\langle b^\dagger(t)b(t')\rangle_{\rm ss} &=&
\sum_{\sigma=\pm} e^{\lambda^*_\sigma \tau}
\left[\sigma\frac{iG}{2\delta}\mathcal{C}_{ab}
+\frac{1}{2}\left(1+\sigma\frac{\Gamma}{\delta}\right)n_b
\right], \nonumber \\
\langle a^\dagger(t)b(t')\rangle_{\rm ss} &=&
\sum_{\sigma=\pm} e^{\lambda^*_\sigma \tau}
\left[\frac{1}{2}\left(1-\sigma\frac{\Gamma}{\delta}\right)\mathcal{C}_{ab}
+\sigma \frac{iG}{2\delta}n_b
\right], \nonumber \\
\langle b^\dagger(t)a(t')\rangle_{\rm ss}
&=&
\sum_{\sigma=\pm} e^{\lambda^*_\sigma \tau}
\left[
\sigma\frac{iG}{2\delta}n_a
+\frac{1}{2}\left(1+\sigma\frac{\Gamma}{\delta}\right)\mathcal{C}_{ba}
\right], \nonumber \label{final_correlations} \\
\end{eqnarray}
where $\Gamma = (\kappa - \gamma)/4 > 0$ and $\delta = \sqrt{\Gamma^2 - G^2}$. In Eq. (\ref{final_correlations}), the square root $\delta=\sqrt{\Gamma^2-G^2}$ is understood with the branch convention consistent with the eigenvalue labels of $M_{ab}^*$; for $G>\Gamma$, replacing $\delta$ by $\delta^*$ merely relabels $\sigma=\pm$ and leaves the correlation functions unchanged. Because ${\rm Re}[\lambda_\pm] < 0$, the correlation functions exponentially decay as $\tau \rightarrow \infty$. The steady-state correlation functions derived above exhibit a universal decomposition into two damped exponentials governed by the drift eigenvalues $\lambda_\pm$, reflecting the presence of two hybrid optomechanical normal modes. The parameter $\delta=\sqrt{\Gamma^2-G^2}$ controls the qualitative dynamical behavior: for $G<\Gamma$ the correlations decay in an overdamped manner with two distinct relaxation rates, while for $G>\Gamma$ they display oscillatory exchange with a normal-mode frequency splitting. While the expressions quoted above assume $\delta \neq 0$, the LEP occurs at $\delta = 0$ where both the roots coalesce and the dynamics exhibits polynomial factors in $\tau$, a feature that is well known at exceptional points \cite{Cartarius_2011}. 

\vspace{2mm}

The (co)variances appearing in $\mathcal{V}$ that determine the correlation functions above can be found straightforwardly to give (see Appendix (\ref{appB}))
\begin{eqnarray}
n_a &=& \frac{4G^2\gamma}{(\kappa+\gamma)(\kappa\gamma+4G^2)}n_{\rm th}, \label{n_a_var} \\
n_b &=& \frac{\gamma\big(\kappa(\kappa+\gamma)+4G^2\big)}{(\kappa+\gamma)(\kappa\gamma+4G^2)}n_{\rm th}, \label{n_b_var} \\
\mathcal{C}_{ba}
&=& -\frac{2i\kappa G\gamma}{(\kappa+\gamma)(\kappa\gamma+4G^2)}n_{\rm th}, \label{ab_cov}
\end{eqnarray} with $\mathcal{C}_{ab}=\mathcal{C}_{ba}^*$. The steady-state (co)variances demonstrate sideband cooling of the mechanics by the effectively zero-temperature optical bath, yielding $n_b<n_{\rm th}$ for any finite coupling and approaching $n_b \simeq \frac{\gamma}{(\kappa+\gamma)}n_{\rm th}$ in the strong-coupling regime. The cavity fluctuations acquire a finite occupation $n_a \propto n_{\rm th}$ through thermal noise from the mechanical mode. Altogether, the correlations and covariances provide the Gaussian characterization relevant to the red-sideband beam-splitter dynamics considered here. 

\section{Liouvillian versus Hamiltonian exceptional points}\label{LvsH_sec}
It is well known that the Lindblad equation leads to an effective non-Hermitian Hamiltonian dictating the conditional no-jump dynamics \cite{Dalibard_1992}. For cavity optomechanics at the red sideband, we shall now show that the exceptional point arising from this non-Hermitian Hamiltonian differs from that obtained from the Liouvillian drift matrix; see \cite{Kopciuch_2025} for a similar analysis for atomic vapors. An important distinction from \cite{Kopciuch_2025} is that while their atomic-vapor model is formulated in a finite-dimensional Hilbert space of internal levels, the optomechanical system considered here involves two bosonic modes with infinite-dimensional Fock spaces for each mode, despite exact solvability in the Gaussian sector. This shall allow us to clearly distinguish the HEP from its Liouvillian counterpart.

\vspace{2mm}

To this end, let us recall that the effective non-Hermitian Hamiltonian for the no-jump dynamics of the dissipative system is given by the well-known expression
\begin{equation}\label{NH}
H_{\rm NH} = H - \frac{i}{2} \sum_k L_k^\dagger L_k,
\end{equation} for the Lindblad operators $\{L_k\}$. For the present situation we have three such operators corresponding to photon decay from the cavity, phonon absorption from the thermal bath, and phonon decay into the bath. These are given, respectively, by 
\begin{eqnarray}
L_a &=& \sqrt{\kappa} a, \label{La_def}\\
L_+ &=& \sqrt{\gamma n_{\rm th}} b^\dagger, \label{L+_def}\\
L_- &=& \sqrt{\gamma(n_{\rm th} + 1)}b. \label{L-_def}
\end{eqnarray}
Since the photon bath is taken to be at zero temperature, there is no analogous photon creation process. Using these operators, the effective non-Hermitian Hamiltonian [Eq. (\ref{NH})] takes the following simple form:
\begin{eqnarray}
H_{\rm NH} &=& -\Delta a^\dagger a + \omega_m b^\dagger b + G(a b^\dagger + a^\dagger b) \\
&& -\frac{i}{2}\kappa a^\dagger a -\frac{i}{2}\gamma(2n_{\rm th}+1) b^\dagger b, \nonumber
\end{eqnarray} up to a constant. One can thus immediately identify that the conditional no-jump single-particle evolution is determined by the matrix
\begin{equation}\label{M_NH_def}
M_{\rm NH} =
\begin{pmatrix}
i\Delta - \frac{\kappa}{2} & -iG\\
-iG & -i\omega_m - \frac{\gamma}{2} (2n_{\rm th} + 1)
\end{pmatrix}.
\end{equation}
More precisely, $M_{\rm NH}$ is the non-Hermitian single-particle mode matrix associated with the quadratic no-jump Hamiltonian $H_{\rm NH}$, rather than an ordinary Hermitian-Heisenberg drift. The eigenvalues are
\begin{equation}\label{lambdaNH_pm}
\lambda_{{\rm NH},\pm} = \frac{1}{2}\Big(i(\Delta-\omega_m)-\frac{\kappa+\gamma(2n_{\rm th} + 1)}{2}\Big) \pm \sqrt{D_{\rm NH}},
\end{equation} with
\begin{equation}
D_{\rm NH} = \frac{1}{4}\Big(i(\Delta+\omega_m)-\frac{\kappa-\gamma (2n_{\rm th} + 1)}{2}\Big)^2 -G^2.
\end{equation}
The supermodes thus coalesce at the red-sideband point ($\Delta = -\omega_m$) if
\begin{equation}\label{G_HEP}
G_{\rm HEP} = \frac{|\kappa - \gamma_{\rm eff}|}{4}, \quad \quad \gamma_{\rm eff} = \gamma (2n_{\rm th} + 1), 
\end{equation} in which case, $M_{\rm NH}$ can be verified to be defective. One notes a thermal damping $\gamma_{\rm eff}$ arising in the conditional no-jump dynamics described by the effective non-Hermitian Hamiltonian $H_{\rm NH}$. At finite temperature, $n_{\rm th} > 0$, and the HEP occurring at $G_{\rm HEP}$ differs from the LEP at $G_{\rm LEP}$ [Eq. (\ref{G_EP})]. This effect, due to thermal phonons, can be interpreted as the mechanical analog of the shift between the LEP and the HEP discussed in \cite{Chimczak_2023} due to the presence of thermal photons in a microwave-frequency system. Notably, if the phonon bath is at zero temperature, i.e., $n_{\rm th} = 0$, then $G_{\rm LEP} = G_{\rm HEP}$.

\vspace{2mm}

The appearance of the thermal dependence in the HEP has a clear physical interpretation. The effective non-Hermitian Hamiltonian $H_{\rm NH}$ governs the conditional evolution of the system under continuous monitoring, i.e., the dynamics is conditioned on the absence of quantum jumps. At finite temperature of the mechanical bath, the Lindblad structure contains not only phonon-loss processes ($b$-jumps) but also phonon-gain processes ($b^\dagger$-jumps), the latter corresponding to thermal absorption from the bath. Even when no jump occurs, the mere possibility of these absorption processes contributes to the imaginary part of $H_{\rm NH}$ through the term $-\frac{i}{2}L_+^\dagger L_+$, effectively enhancing the conditional damping rate of the mechanical mode to $\gamma_{\rm eff}=\gamma(2n_{\rm th}+1)$. As a consequence, the coalescence condition for the conditional normal modes, i.e., $G =G_{\rm HEP}$, becomes explicitly temperature dependent. By contrast, the LEP reflects the unconditional Lindbladian evolution in which thermal excitation and relaxation channels enter symmetrically and do not renormalize the drift matrix. The thermal shift of the HEP therefore highlights the fundamentally different nature of the conditional no-jump dynamics versus the complete Liouvillian description at finite temperature.

\vspace{2mm}

\begin{figure}[htbp]
\centering

\subfloat[]{
  \includegraphics[width=\columnwidth]{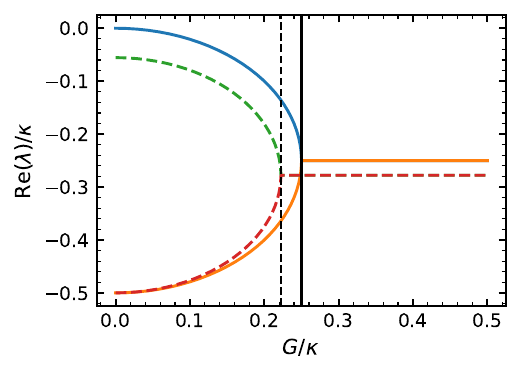}
  \label{fig:left}
}\\[2mm]
\subfloat[]{
  \includegraphics[width=\columnwidth]{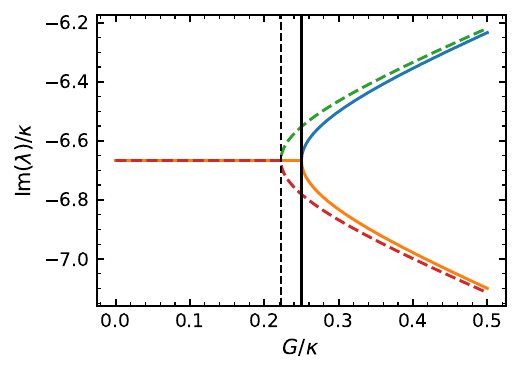}
  \label{fig:right}
}

\caption{\justifying{Behavior of the eigenvalues across the exceptional point: (a) real parts and (b) imaginary parts. Solid curves show the eigenvalue branches $\lambda_{\pm}$ while dashed curves correspond to the eigenvalue branches $\lambda_{{\rm NH},\pm}$. The vertical black lines mark the locations of the exceptional points; solid for LEP and dashed for HEP. The parameters used are $\kappa/2\pi=150~{\rm kHz}$,
$\gamma/2\pi=0.1~{\rm Hz}$,
$\omega_m/2\pi=1~{\rm MHz}$,
and thermal occupation $n_{\rm th}=8.33\times10^{4}$.}}
\label{fig1}
\end{figure}

In Fig. (\ref{fig1}), we have plotted the eigenvalues $\lambda_\pm$ [Eq. (\ref{lambda_pm})] from the Liouvillian drift matrix and those [Eq. (\ref{lambdaNH_pm})] from the no-jump Hamiltonian for a sample set of physical parameters $\omega_m/2\pi = 1~{\rm MHz}$, $\gamma/2\pi = 0.1~{\rm Hz}$, $\kappa/2\pi = 150~{\rm kHz}$ \cite{Underwood_2015}, and $n_{\rm th} \approx 8.33 \times 10^4$ for a cryogenic Helium bath at $T = 4~{\rm K}$. For these parameters, the LEP and HEP occur at
\begin{equation}
G_{\rm LEP} \approx 0.25 \kappa, \quad \quad G_{\rm HEP} \approx 0.22 \kappa,
\end{equation}
respectively, as can be observed from Fig. (\ref{fig1}). Since $G$ is controlled by the input power of the control drive about which linearization has been performed as $G \propto \sqrt{P_{\rm in}}$, 
the corresponding change in the required input power is
\begin{equation}
\frac{P_{\rm in,HEP}}{P_{\rm in,LEP}} = \left(\frac{G_{\rm HEP}}{G_{\rm LEP}}\right)^2
\approx \left(\frac{0.22}{0.25}\right)^2 \approx 0.8,
\end{equation}
implying that the transition from the LEP to the HEP corresponds to approximately a $20\%$ change in the input power. Thus the separation between the LEP and HEP translates into a significant shift in the control-drive power at which eigenvalue coalescence occurs, suggesting an operational way to distinguish the two notions of exceptional points at finite temperature: thermal absorption processes renormalize the conditional damping entering $H_{\rm NH}$, while the unconditional drift governing the steady-state correlations remains independent of temperature.

\section{Thermofield formalism and hybrid exceptional points}\label{hybrid_sec}
In this section, we will describe hybrid exceptional points that are intermediate between the LEPs and HEPs. This necessitates the introduction of a hybrid-Liouvillian superoperator that continuously interpolates between the Liouvillian and Hamiltonian regimes \cite{Minganti_2020,Kopciuch_2025}. The appropriate and elegant framework to achieve this is the thermofield formalism \cite{Takahashi_1996,Jamiolkowski_1972,Choi_1975} in which the density matrix $\rho$ undergoes the vectorization 
\begin{equation}\label{rhoket_def}
|\rho\rangle = \sum_{ij} \rho_{ij} |i\rangle \otimes |\tilde{j}\rangle,
\end{equation} as a state in the doubled Hilbert space $\mathcal{H}\otimes\tilde{\mathcal{H}}$, where $\mathcal{H}$ is the physical Hilbert space and $\tilde{\mathcal{H}}$ is the auxiliary or simply the `tilde' space. The master equation $\frac{d\rho}{dt} = \mathcal{L}(\rho)$ can then be written as \cite{Takahashi_1996} (see \cite{Jamiolkowski_1972,Choi_1975} for the conventions)
\begin{equation}\label{TF_eqn}
\frac{d}{dt}|\rho\rangle = -iH_{\rm TF}|\rho\rangle,
\end{equation}
with a non-Hermitian thermofield Hamiltonian $H_{\rm TF} = i \mathcal{L}$. A Lindblad master equation is thus mapped to a Schr\"odinger-type equation. 

\subsection{Thermofield equation for red-sideband optomechanics}\label{thermo_sec}
To formulate the thermofield dynamics for the red-sideband physics, let us introduce a doubled Hilbert space $\mathcal{H} \otimes \tilde{\mathcal{H}}$ and map the density matrix to a vector in the manner made explicit in Eq. (\ref{rhoket_def}). Left and right multiplication act as \cite{Jamiolkowski_1972,Choi_1975}
\begin{equation}
O\rho  \mapsto (O\otimes\mathbb{I}) |\rho\rangle,\quad \quad
\rho O  \mapsto (\mathbb{I}\otimes O^{T}) |\rho\rangle,
\end{equation}
where tilde operators satisfy the same bosonic commutation relations as the physical ones. In the Fock basis, one has $a^{T}=a^\dagger$ and $(a^\dagger)^{T}=a$ (similarly for $b$), meaning that right multiplication can be written in terms of tilde operators as
\begin{equation}
\rho a \mapsto \tilde{a}^\dagger|\rho\rangle,\quad
\rho a^\dagger \mapsto \tilde{a}|\rho\rangle,\quad
\rho b \mapsto \tilde{b}^\dagger|\rho\rangle,\quad
\rho b^\dagger \mapsto \tilde{b}|\rho\rangle,
\end{equation} and the master equation is cast in the form given in Eq. (\ref{TF_eqn}) whose Hamiltonian we shall now find. The conservative Hamiltonian contribution $-i[H,\rho]$ gives the following expression on the doubled space:
\begin{equation}
H_{\rm TF}^{(H)} = H \otimes \mathbb{I} - \mathbb{I} \otimes \tilde{H},
\end{equation} where $\tilde{H} = H^T$ acting on the auxiliary Hilbert space $\tilde{\mathcal{H}}$. 
Explicitly one has
\begin{eqnarray}
H_{\rm TF}^{(H)} &=& -\Delta (a^\dagger a - \tilde{a}^\dagger \tilde{a}) + \omega_m (b^\dagger b - \tilde{b}^\dagger \tilde{b}) \label{H_TF_H} \\
&&+ G\left[(a^\dagger b+ab^\dagger) - (\tilde{a}^\dagger \tilde{b}+\tilde{a}\tilde{b}^\dagger)\right]. \nonumber
\end{eqnarray}
For the dissipative part, let us note that each Lindblad term generically contributes $\mathcal{D}[o]\rho \to \left(o\tilde{o} - \frac{1}{2}o^\dagger o - \frac{1}{2}\tilde{o}^\dagger \tilde{o}\right)$, thereby implying that the full thermofield Hamiltonian goes as
\begin{eqnarray}
H_{\rm TF} &=& H_{\rm TF}^{(H)}
+ i\kappa\left(a\tilde{a} - \frac{1}{2}a^\dagger a - \frac{1}{2}\tilde{a}^\dagger \tilde{a}\right) \label{HTF_RSB} \\
&&+ i\gamma (n_{\rm th}+1)\left(b\tilde{b} - \frac{1}{2}b^\dagger b - \frac{1}{2}\tilde{b}^\dagger \tilde{b}\right) \nonumber \\
&&+ i\gamma n_{\rm th}\left(b^\dagger \tilde{b}^\dagger - \frac{1}{2}b b^\dagger - \frac{1}{2}\tilde{b} \tilde{b}^\dagger\right), \nonumber
\end{eqnarray}
with $H_{\rm TF}^{(H)}$ given by Eq. (\ref{H_TF_H}). This non-Hermitian Hamiltonian fully encodes the optomechanical Lindblad dynamics near the red sideband and also makes explicit a continuous symmetry (see Appendix (\ref{appC})). 

\subsection{Liouvillian exceptional points}
As a natural precursor to utilizing the thermofield formalism to probe hybrid exceptional points, let us take a brief detour to demonstrate how the LEP arises from this framework in a natural manner. Since the thermofield Hamiltonian is quadratic in the canonical operators, the dynamics of the operator vector $\Phi= (a, ~b, ~\tilde{a}^\dagger, ~\tilde{b}^\dagger)^T$ closes exactly as
\begin{equation}
\frac{d\Phi}{dt}=i\left[H_{\rm TF},\Phi\right]
=M \Phi,
\label{structure_eom_final}
\end{equation}
where $M$ encodes the linearized action of the thermofield generator on the left and right operators rather than the physical Heisenberg drift. At the exact red-sideband detuning $\Delta=-\omega_m$ and up to a shift, one finds the following $4\times4$ matrix:
\begin{equation}
M=
\begin{pmatrix}
-\frac{\kappa}{2} & -iG & 0 & 0 \\
-iG & -\frac{\gamma (2n_{\rm th}+1)}{2} & 0 & \gamma n_{\rm th} \\
-\kappa & 0 & \frac{\kappa}{2} & -iG \\
0 & -\gamma(n_{\rm th}+1) & -iG & \frac{\gamma(2n_{\rm th}+1)}{2}
\end{pmatrix},
\label{M_matrix}
\end{equation}
with the nonzero off-block-diagonal elements indicating the mixing between the physical and auxiliary sectors in the thermofield space. This mixing arises due to the terms $i\kappa a \tilde{a}$, $i\gamma(n_{\rm th} + 1)b \tilde{b}$, and $i \gamma n_{\rm th} b^\dagger \tilde{b}^\dagger$ that the thermofield Hamiltonian [Eq. (\ref{HTF_RSB})] contains. Quite remarkably, despite the complicated structure, the characteristic equation of $M$ admits the simple expression
\begin{equation}\label{M_charac}
\lambda^4 - \left[ \frac{\kappa^2 + \gamma^2}{4} - 2G^2\right] \lambda^2 + \left[G^4 + \frac{\kappa \gamma}{2} G^2 + \left(\frac{\kappa \gamma}{4}\right)^2 \right]=0.
\end{equation}
Defining $A = \frac{\kappa^2 + \gamma^2}{4} - 2G^2$ and $B = G^4 + \frac{\kappa \gamma}{2} G^2 + \left(\frac{\kappa \gamma}{4}\right)^2$, an exceptional point occurs when $A^2 - 4B = 0$. This algebraic condition can be solved to precisely yield the expression $G = G_{\rm LEP}$ [Eq. (\ref{G_EP})], thereby showing that the LEP as identified earlier from the singularity of the physical drift is precisely the one appearing in the thermofield space where the physical and auxiliary degrees of freedom mix. It should be clarified here that although Eq. (\ref{M_charac}) is quartic in $\lambda$, it is quadratic in $s=\lambda^2$, i.e., if its two roots are $s_1$ and $s_2$, then the four roots in $\lambda$ are $\lambda=\pm\sqrt{s_1}$ and $\lambda=\pm\sqrt{s_2}$. At the exceptional point $s_1=s_2$, giving two symmetric pairwise coalescences $\sqrt{s_1}=\sqrt{s_2}$ and $-\sqrt{s_1}=-\sqrt{s_2}$, rather than a single fourfold coalescence. One therefore has second-order exceptional-point degeneracies, not a higher-order exceptional point. Apart from reproducing Eq. (\ref{G_EP}), the thermofield approach offers a tantalizing insight: while the thermofield mixing involves the thermal Bose factor $n_{\rm th}$ as explicit in Eq. (\ref{M_matrix}), the characteristic equation [Eq. (\ref{M_charac})] that governs the spectrum of the time evolutions of the operators turns out to be independent of $n_{\rm th}$, i.e., the Bose factor undergoes a precise cancellation. This is in complete agreement with the observation that the drift matrix $M_{ab}$ is independent of $n_{\rm th}$ and the thermal factor only enters at the level of the two-point functions, consistent with the use of noise correlators in the Langevin framework. 

\subsection{Hybrid-Liouvillian superoperator and exceptional points}
While we already observed the distinction between the LEP and the HEP, let us introduce a hybrid-Liouvillian superoperator \cite{Minganti_2020,Kopciuch_2025} whose spectrum continuously interpolates between the two regimes. Introducing a real parameter $0 \leq \epsilon \leq 1$, the hybrid-Liouvillian superoperator is defined as 
\begin{eqnarray}
\mathcal{L}_\epsilon(\rho) &=& -i\Big(H_{\rm NH}\rho-\rho H_{\rm NH}^\dagger\Big) \label{hybrid_Liouville} \\
&& +\epsilon\Big(
\kappa a\rho a^\dagger
+\gamma(n_{\rm th}+1) b\rho b^\dagger
+\gamma n_{\rm th} b^\dagger\rho b
\Big), \nonumber
\end{eqnarray}
where $H_{\rm NH}$ is given in Eq. (\ref{NH}). So $\epsilon = 0$ corresponds to the Hamiltonian no-jump dynamics dictated by $H_{\rm NH}$ while $\epsilon = 1$ corresponds to the full Liouvillian dynamics. In other words, $\epsilon$ signifies the strength of the quantum jumps in the dissipative optomechanical dynamics. Thus by dialing the parameter $\epsilon$ one can probe hybrid exceptional points lying between the Hamiltonian and Liouvillian regimes. The parameter $\epsilon$ can be referred to as the quantum-jump parameter
\cite{Minganti_2020,Kopciuch_2025}. 

\vspace{2mm}

\begin{figure}
    \centering
    \includegraphics[width=1\linewidth]{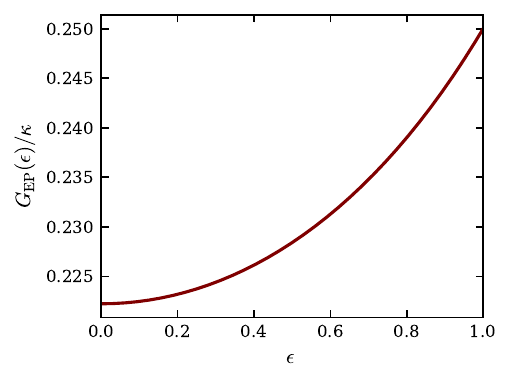}
    \caption{\justifying{Variation of $G_{\rm EP}(\epsilon)$ between the HEP ($\epsilon=0$) and LEP ($\epsilon=1$) limits. The parameters used are $\kappa/2\pi=150~{\rm kHz}$, $\gamma/2\pi=0.1~{\rm Hz}$, and thermal occupation $n_{\rm th}=8.33\times10^{4}$.}}
\label{fig2}
\end{figure}

Given a generic operator $O$, the adjoint hybrid-Liouvillian evolution is given by
\begin{equation}
\mathcal{L}_\epsilon^\dagger(O) = i\Big(H_{\rm NH}^\dagger O-OH_{\rm NH}\Big) +\epsilon\sum_{k} L_k^\dagger O L_k,
\label{adjoint_hybrid}
\end{equation} with the jump operators given in Eqs. (\ref{La_def}), (\ref{L+_def}), and (\ref{L-_def}). The Liouvillian dynamics thus encodes only partial or weighted quantum jumps for $0 < \epsilon < 1$. The quantum master equation dictated by the hybrid-Liouvillian $\mathcal{L}_\epsilon$ can be analyzed in the thermofield approach on the doubled Hilbert space $\mathcal{H} \otimes \tilde{\mathcal{H}}$ using the same techniques as adopted in writing Eq. (\ref{HTF_RSB}). Within the thermofield construction, $\frac{d\rho}{dt}=\mathcal{L}_\epsilon(\rho)$ is
mapped to the Schr\"odinger-type equation
$\frac{d}{dt}|\rho\rangle=-iH_{{\rm TF},\epsilon}|\rho\rangle$ with the hybrid-thermofield Hamiltonian
\begin{eqnarray}
H_{{\rm TF},\epsilon}
&=&
H_{\rm NH}\otimes\mathbb I
-\mathbb I\otimes\tilde{H}_{\rm NH}^\dagger \label{hybrid_TF_compact} \\
&& +i\epsilon\Big(
\kappa a\tilde{a}
+\gamma(n_{\rm th}+1) b\tilde{b}
+\gamma n_{\rm th} b^\dagger\tilde{b}^\dagger
\Big), \nonumber
\end{eqnarray}
where $\tilde{H}_{\rm NH}^\dagger = (H_{\rm NH}^\dagger)^T$ acts on the auxiliary Hilbert space $\tilde{\mathcal{H}}$. Restricting to the red-sideband regime of interest where $H_{\rm NH}=H-\frac{i}{2}\kappa a^\dagger a
-\frac{i}{2}\gamma(2n_{\rm th}+1)b^\dagger b$, the hybrid-thermofield Hamiltonian admits the following explicit form:
\begin{eqnarray}
H_{{\rm TF},\epsilon}
&=&
-\Delta\big(a^\dagger a-\tilde{a}^\dagger\tilde{a}\big)
+\omega_m\big(b^\dagger b - \tilde{b}^\dagger \tilde{b}\big)\label{hybrid_TF_explicit} \\
&& +G\big(a^\dagger b + a b^\dagger - \tilde{a}^\dagger \tilde{b} - \tilde{a} \tilde{b}^\dagger\big) \nonumber\\
&&
-\frac{i}{2}\kappa\big(a^\dagger a + \tilde{a}^\dagger \tilde{a}\big) -\frac{i}{2}\gamma(2n_{\rm th}+1)\big(b^\dagger b + \tilde{b}^\dagger \tilde{b}\big)
\nonumber\\
&&
+i\epsilon\left(\kappa a \tilde{a} + \gamma(n_{\rm th}+1) b \tilde{b} + \gamma n_{\rm th} b^\dagger \tilde{b}^\dagger \right), \nonumber
\end{eqnarray}
up to an additive constant. One can verify immediately that $\epsilon=1$ gives the thermofield Hamiltonian $H_{\rm TF}$ given in Eq. (\ref{HTF_RSB}). For a given $0 \leq \epsilon \leq 1$, the thermofield-space dynamics of the operator vector $\Phi = (a, ~b, ~\tilde{a}^\dagger, ~\tilde{b}^\dagger)$ is described by the form given in Eq. (\ref{structure_eom_final}), but with the modified $\epsilon$-dependent matrix
\begin{equation}
M_\epsilon=
\begin{pmatrix}
-\frac{\kappa}{2} & -iG & 0 & 0 \\
-iG & -\frac{\gamma (2n_{\rm th}+1)}{2} & 0 & \epsilon\gamma n_{\rm th} \\
-\epsilon\kappa & 0 & \frac{\kappa}{2} & -iG \\
0 & -\epsilon\gamma(n_{\rm th}+1) & -iG & \frac{\gamma(2n_{\rm th}+1)}{2}
\end{pmatrix}.
\label{M_epsilon}
\end{equation}
While $M_\epsilon$ reduces to Eq. (\ref{M_matrix}) when $\epsilon = 1$, reproducing the Liouvillian result, it simplifies significantly if one considers the conditional no-jump dynamics, i.e., the Hamiltonian regime ($\epsilon = 0$). Putting $\epsilon = 0$ reveals a block-diagonal structure, meaning that the physical and tilde sectors completely decouple in the no-jump regime. In that case, the upper-diagonal block coincides with $M_{\rm NH}$ [Eq. (\ref{M_NH_def})] up to the uniform shift $\sim -i\Delta \mathbb{I} = +i\omega_m \mathbb{I}$ at the red-sideband point; this naturally leads to the same result for the HEP as derived in Eq. (\ref{G_HEP}), confirming that the hybrid-Liouvillian description in the thermofield space is consistent with both the $\epsilon = 0,1$ limits. In general, the characteristic polynomial is $\det(M_\epsilon-\lambda\mathbb I) = \lambda^4-\alpha\lambda^2+\beta$, where
\begin{widetext}
\begin{eqnarray}
\alpha &=& \frac{1}{4}\Big[\kappa^2+\gamma^2
+4\gamma^2n_{\rm th}(n_{\rm th}+1)(1-\epsilon^2)\Big]
-2G^2, \\
\beta &=& G^4+\Big(\frac{\kappa\gamma}{2}
+n_{\rm th}\kappa\gamma(1-\epsilon^2)\Big)G^2 +\frac{\kappa^2\gamma^2}{16} \left[1+4n_{\rm th}(n_{\rm th}+1)(1-\epsilon^2)\right].
\end{eqnarray}
\end{widetext}
A hybrid exceptional point occurs when this quartic possesses a repeated root, i.e., when the discriminant of the quadratic in $s=\lambda^2$ vanishes, i.e., $\alpha^2-4\beta=0$. Solving this condition yields an explicit analytic expression for the hybrid exceptional-point location:
\begin{equation}
G_{\rm EP}(\epsilon) = \frac{
\left|\kappa^2-\gamma^2\left[1+4n_{\rm th}(n_{\rm th}+1)(1-\epsilon^2)\right]\right|}{4\sqrt{(\kappa+\gamma)^2+4n_{\rm th}\gamma\left[\kappa+\gamma(n_{\rm th}+1)\right](1-\epsilon^2)}}.
\label{G_EP_hybrid}
\end{equation}
This result interpolates continuously between the two limiting exceptional points, i.e., $G_{\rm EP}(0)
= G_{\rm HEP} = \frac{|\kappa-\gamma(2n_{\rm th}+1)|}{4}$ [Eq. (\ref{G_HEP})] and $G_{\rm EP}(1) = G_{\rm LEP} = \frac{\kappa-\gamma}{4}$ [Eq. (\ref{G_EP})]. The thermofield approach to the hybrid-Liouvillian evolution thus accurately captures the exceptional points in situations with partial quantum jumps, i.e., intermediate between the Liouvillian and no-jump descriptions. In Fig. (\ref{fig2}), we have demonstrated the functional dependence of $G_{\rm EP}(\epsilon)$ on $\epsilon$, showing a nonlinear (quadratic) dependence. The latter result has a profound implication: if one considers the effect of weak quantum jumps, i.e., $\epsilon \approx 0$ near the conditional no-jump regime, the location of the HEP acquires no first-order or linear corrections in $\epsilon$, indicating the robustness of the HEP against fluctuations in $\epsilon$ about $\epsilon = 0$. 

\vspace{2mm}

From a physical viewpoint, the interpolation parameter $\epsilon$ should be interpreted as a spectral deformation of the dynamical generator that continuously tunes the relative weight of quantum-jump terms against the non-Hermitian no-jump evolution. In particular, for $\epsilon=0$ the dynamics is governed solely by the effective non-Hermitian Hamiltonian $H_{\rm NH}$ and corresponds to the unnormalized no-jump evolution (i.e., evolution conditioned on the absence of all detection events), which at finite temperature leads to the enhanced conditional damping $\gamma_{\rm eff}=\gamma(2n_{\rm th}+1)$ and hence to the HEP. For $\epsilon=1$ one recovers the full trace-preserving Lindblad evolution in which the jump and no-jump contributions combine to yield the unconditional Liouvillian dynamics and hence the LEP. For intermediate values $0<\epsilon<1$, the superoperator $\mathcal{L}_\epsilon$ is generally not trace preserving and thus should not be viewed as the unconditional master equation for an inefficiently-monitored bath; rather, it defines an unnormalized hybrid evolution that biases the ensemble of trajectories by partially suppressing jump events. In this precise sense, $\epsilon$ provides a continuous interpolation of the spectrum between the Hamiltonian (no-jump) and Liouvillian (unconditional) limits, yielding hybrid exceptional points $G_{\rm EP}(\epsilon)$ that smoothly connect the HEP and the LEP.

\vspace{2mm}

From the experimental viewpoint, the continuous tuning of the quantum-jump parameter $\epsilon$ can possibly be realized through the framework of continuous weak measurement and the postselection of quantum trajectories \cite{Minganti_2020,Wiseman_2009}. By monitoring the relevant dissipation channels and by selectively discarding experimental records that contain specific jump signatures, one can bias the ensemble evolution away from the unconditional Liouvillian evolution ($\epsilon=1$) to the conditional Hamiltonian evolution ($\epsilon=0$). Measurement-conditioned non-Hermitian dynamics and exceptional-point crossings have already been successfully demonstrated using postselection in superconducting circuits \cite{Naghiloo_2019}. Translating these techniques to the state-of-the-art cavity setups which already feature highly-efficient measurement-based quantum control \cite{Rossi_2018} therefore provides a viable pathway for observing these hybrid exceptional points.

\section{Conclusions}\label{conc_sec}
This paper exposes a clear distinction between LEPs and HEPs for optomechanical systems at the red-sideband detuning. Although exceptional points have been frequently described from effective non-Hermitian Hamiltonians \cite{Heiss_2012}, for open systems such an approach disregards quantum jumps that are only taken into account if one probes the spectrum of the Liouvillian \cite{Minganti_2019,Minganti_2020,Sun_2024,Abo_2024,Kopciuch_2025}. The present work shows how cavity optomechanics with a thermal phonon bath offers an elegant platform to explore the role of quantum jumps leading to distinct notions of LEPs and HEPs. In particular, it was shown how the presence of the thermal phonon bath leads to a conditional effective damping, thereby shifting the HEP from the LEP. The physical mechanism behind this shift was traced back to phonon-absorption jumps in which the system absorbs excitations from the phonon bath at nonzero temperature. It should be noted that this shift yields information about the thermal bath and thus may be used as a probe of the thermal phonon number.

\vspace{2mm}

Beyond clarifying the distinction between the LEP and the HEP, we studied a hybrid interpolation between these two regimes via a continuous quantum-jump parameter \cite{Minganti_2020,Kopciuch_2025} $0 \leq \epsilon \leq 1$ whose role is to weigh the effect of quantum jumps. While $\epsilon=0$ and $\epsilon=1$ correspond, respectively, to no quantum jumps and full quantum jumps, its intermediate values give rise to taking into account partial quantum jumps. In the weak-quantum-jump regime, i.e., for $\epsilon \approx 0$, we found that $G_{\rm EP}(\epsilon) \approx G_{\rm HEP} + \mathcal{O}(\epsilon^2)$, showing that the location of the HEP is robust against small hybrid perturbations. Our approach for tackling this hybrid-Liouvillian problem has been based on the thermofield construction, allowing us to derive an exact analytical result for the location of the hybrid exceptional points. Overall, our results demonstrate that the versatile platform of cavity optomechanics can be used to delineate which distinct classes of exceptional points arise depending on whether one considers unconditional Liouvillian evolution, conditional no-jump dynamics, or intermediate hybrid deformations. In this context, the thermofield formalism is a unifying analytical tool to capture these singularities within a single spectral framework, thereby clarifying the physical meaning of LEPs, HEPs, and their continuous hybrid deformations. The versatility and experimental feasibility of cavity platforms suggests the possibility of experimental realization of probing partial quantum jumps and hybrid exceptional points. \\

\textbf{Acknowledgements:} M.B. thanks the Air Force Office of Scientific Research (AFOSR) (FA9550-23-1-0259) for support.

\appendix

\begin{widetext}

\section{Calculation details of steady-state correlation functions}\label{appA}
At the red-sideband point $\Delta=-\omega_m$, the deterministic drift equation for the operator vector $\Psi=(a, ~b)^T$ reads
\begin{equation}
\frac{d}{dt}\Psi=M_{ab} \Psi,
\quad \quad
M_{ab}=
\begin{pmatrix}
-i\omega_m-\kappa/2 & -iG\\
-iG & -i\omega_m-\gamma/2
\end{pmatrix}.
\label{Mab_RSB}
\end{equation}
It is convenient to introduce the parameters
\begin{equation}
\mu=\frac{\kappa+\gamma}{4},
\quad \quad
\Gamma=\frac{\kappa-\gamma}{4},
\end{equation}
so that $M_{ab}$ can be written as
\begin{equation}
M_{ab} = -\big(i\omega_m+\mu\big)\mathbb{I} +
\begin{pmatrix}
-\Gamma & -iG\\
-iG & +\Gamma
\end{pmatrix}.
\label{Mab_split}
\end{equation}
The drift matrix possesses two eigenvalues that go as
\begin{equation}
\lambda_\pm=-i\omega_m-\mu\pm\delta, \quad \quad \delta=\sqrt{\Gamma^2-G^2}.
\label{lambda_pm_app}
\end{equation}
The parameter $\delta$ controls the qualitative behavior of the coupled system: for $G<\Gamma$ it is real, while for $G>\Gamma$ it is purely imaginary. The LEP occurs at $\delta=0$.

\vspace{2mm}

For the normal-ordered correlation functions, the relevant dynamics is described by the matrix $M_{ab}^*$ which reads
\begin{equation}
M_{ab}^*=
\begin{pmatrix}
+i\omega_m-\kappa/2 & +iG\\
+iG & +i\omega_m-\gamma/2
\end{pmatrix}.
\label{Mab_star_app}
\end{equation}
The eigenvalues of $M_{ab}^*$ are simply $\lambda_\pm^*$, which are
\begin{equation}
\lambda_\pm^*=+i\omega_m-\mu\pm\delta^*,
\label{lambda_star_pm_app}
\end{equation}
where $\delta^*=\delta$ if $G < \Gamma$ and $\delta^* = -\delta$ if $G > \Gamma$. The time-evolution operator therefore admits the spectral expansion given in Eq. (\ref{spectral_expansion}). Using the eigenvalue splitting
\begin{equation}
\lambda_+^*-\lambda_-^*=2\delta^*,
\end{equation}
and substituting Eq. (\ref{Mab_star_app}), one finds
\begin{equation}
P_\pm^* = \frac{1}{2}\left(
\mathbb{I} \pm \frac{1}{\delta^*}
\begin{pmatrix}
-\Gamma & +iG\\
+iG & +\Gamma
\end{pmatrix}
\right).
\label{Ppm_matrix_app}
\end{equation}
Equivalently, the individual matrix elements are
\begin{equation}\label{P_explicit}
(P_\pm^*)_{11} = \frac{1}{2}\left(1\mp\frac{\Gamma}{\delta^*}\right), \quad \quad (P_\pm^*)_{22} =
\frac{1}{2}\left(1\pm\frac{\Gamma}{\delta^*}\right), \quad \quad (P_\pm^*)_{12}=(P_\pm^*)_{21} = \pm\frac{iG}{2\delta^*}.
\end{equation}
These projectors satisfy the standard identities
\begin{equation}
P_+^*+P_-^*=\mathbb{I},
\quad \quad
(P_\pm^*)^2=P_\pm^*,
\quad \quad
P_+^*P_-^*=0.
\end{equation}
Eq. (\ref{Ctau_matrix_form}) implies
\begin{equation}
C(\tau) = e^{M_{ab}^*\tau} \mathcal{V}
= \sum_{\sigma=\pm}
P_\sigma^* \mathcal{V} e^{\lambda_\sigma^*\tau},
\label{Ctau_projector_app}
\end{equation}
which, using Eq. (\ref{P_explicit}) gives rise to the closed-form steady-state correlation functions quoted in Eq. (\ref{final_correlations}) of the main text. It holds for both $G < \Gamma$ and $G > \Gamma$, corresponding to $\delta = \delta^*$ and $\delta^* = -\delta$, respectively, as the latter simply exchanges the labels $\pm$ and leaves all the
correlation functions [Eq. (\ref{final_correlations})] invariant.

\section{Calculation of the steady-state covariances at the exact red-sideband point}\label{appB}
Here one starts with the Heisenberg-Langevin equations in the steady state, so the (co)variances are independent of time, i.e., 
\begin{eqnarray}
0&=&\frac{d}{dt}\langle a^\dagger a\rangle_{\rm ss}
=
-\kappa\langle a^\dagger a\rangle_{\rm ss}
+iG\Big(\langle b^\dagger a\rangle_{\rm ss}-\langle a^\dagger b\rangle_{\rm ss}\Big),
\label{eq_na_TF} \\
0&=& \frac{d}{dt}\langle b^\dagger b\rangle_{\rm ss}
=
-\gamma\langle b^\dagger b\rangle_{\rm ss}
+\gamma n_{\rm th}
-iG\Big(\langle b^\dagger a\rangle_{\rm ss}-\langle a^\dagger b\rangle_{\rm ss}\Big), \label{eq_nb_TF}\\
0&=&\frac{d}{dt}\langle b^\dagger a\rangle_{\rm ss}
=
-\frac{\kappa+\gamma}{2}\langle b^\dagger a\rangle_{\rm ss}
+iG\Big(\langle a^\dagger a\rangle_{\rm ss}-\langle b^\dagger b\rangle_{\rm ss}\Big),
\label{eq_Cba_TF}
\end{eqnarray}
where we have taken $\Delta+\omega_m=0$. Eqs. (\ref{eq_na_TF})-(\ref{eq_Cba_TF}) form a closed linear system for the steady-state second moments. Writing
\begin{equation}
n_a= \langle a^\dagger a\rangle_{\rm ss},\quad
n_b= \langle b^\dagger b\rangle_{\rm ss},\quad
\mathcal{C}_{ba}= \langle b^\dagger a\rangle_{\rm ss},\quad
\mathcal{C}_{ab}= \langle a^\dagger b\rangle_{\rm ss}=\mathcal{C}_{ba}^*,
\end{equation}
the Eqs. (\ref{eq_na_TF}) and (\ref{eq_nb_TF}) imply
\begin{equation}
\kappa n_a=\gamma(n_{\rm th}-n_b).
\label{balance_relation}
\end{equation}
Eq. (\ref{eq_Cba_TF}) gives the coherence in terms of the population imbalance as
\begin{equation}
\mathcal{C}_{ba}
=
-\frac{2iG}{\kappa+\gamma}(n_b-n_a).
\label{Cba_relation}
\end{equation}
Inserting Eq. (\ref{Cba_relation}) into Eq. (\ref{eq_na_TF}) yields
\begin{equation}
\kappa n_a=\frac{4G^2}{\kappa+\gamma}(n_b-n_a),
\label{coupling_relation}
\end{equation}
which together with (\ref{balance_relation}) determines $n_a$ and $n_b$ uniquely. Solving Eqs. (\ref{balance_relation}) and (\ref{coupling_relation}) gives rise to the expressions quoted in Eqs. (\ref{n_a_var}) and (\ref{n_b_var}). The coherence then follows from Eq. (\ref{Cba_relation}) to yield the expression quoted in Eq. (\ref{ab_cov}). 

\section{Charge-sector decomposition}\label{appC}
An immediate insight that emerges from the thermofield construction is that the following is a conserved quantity: 
\begin{equation}
\mathcal{N} = (a^\dagger a + b^\dagger b) - (\tilde{a}^\dagger \tilde{a} + \tilde{b}^\dagger \tilde{b}),
\end{equation} i.e., $[\mathcal{N}, H_{\rm TF}] = 0$, meaning that given an initial state $|\rho(0) \rangle$ at fixed $\mathcal{N}$, its evolution $|\rho(t) \rangle$ under the thermofield dynamics, while nonunitary, conserves $\mathcal{N}$. Now the formal solution of the thermofield equation [Eq. (\ref{TF_eqn})] is given by
\begin{equation}
|\rho(t)\rangle = U(t)|\rho(0)\rangle, \quad \quad U(t)=e^{-itH_{\rm TF}}.
\end{equation}
Let $\mathcal{S}_n=\ker(\mathcal{N}-n \mathbb{I})$ denote the eigenspace of $\mathcal{N}$ with eigenvalues $n\in\mathbb{Z}$. One can then define the projector $P_n$ as
\begin{equation}
P_n =\sum_{\substack{n_a,n_b,\tilde{n}_a,\tilde{n}_b\geq 0\\ n = (n_a + n_b) - (\tilde{n}_a + \tilde{n}_b)}}
\big|n_a,n_b\big\rangle\big\langle n_a,n_b\big|\otimes\big|\tilde{n}_a,\tilde{n}_b\big\rangle\big\langle \tilde{n}_a,\tilde{n}_b\big|.
\label{Pq}
\end{equation}
The commutation relation $[\mathcal{N}, H_{\rm TF}]=0$ implies that $H_{\rm TF}$ is block diagonal with respect to the decomposition $\mathcal{H}\otimes\tilde{\mathcal{H}}=\bigoplus_{n\in\mathbb{Z}}\mathcal{S}_n$, namely, $H_{\rm TF}=\sum_{n\in\mathbb Z} H_n$, and
\begin{equation}
U(t)=e^{-itH_{\rm TF}} =\sum_{n\in\mathbb Z} P_n e^{-it H_n} P_n. \label{U_sector_sum}
\end{equation}
Equivalently, decomposing $|\rho(0)\rangle=\sum_n |\rho^{(n)}(0)\rangle$ with
$|\rho^{(n)}(0)\rangle=P_n|\rho(0)\rangle$, each component evolves independently. In the thermofield-space number basis, this expresses the statement that the density-matrix elements
$\rho_{n_a n_b;\tilde{n}_a\tilde{n}_b}$ with fixed
$n$ do not mix with other orders. 

\end{widetext}

\end{document}